\documentclass[aps,prd,showpacs,showkeys,amsmath,amssymb]{revtex4}
\usepackage{graphicx}
\usepackage{dcolumn}
\usepackage{axodraw}
\usepackage{bm}
\begin{document}

\title{The Possible $J^{PC}I^G=2^{++}2^+$ State X(1600)}

\author{W. Wei, L. Zhang}
\affiliation{%
Department of Physics, Peking University, BEIJING 100871, CHINA}
\author{Shi-Lin Zhu}
\email{zhusl@th.phy.pku.edu.cn} \affiliation{ RCNP, Osaka
University, Japan} \affiliation{ Department of Physics, Peking
University, Beijing 100871, China}

\begin{abstract}

The interesting state X(1600) with $J^{PC}I^G=2^{++}2^+$ can't be
a conventional $q \bar q$ meson in the quark model. Using a mixed
interpolating current with different color configurations, we
investigate the possible existence of X(1600) in the framework of
QCD finite energy sum rules. Our results indicate that both the
"hidden color" and coupled channel effects may be quite important
in the multiquark system. We propose several reactions to look for
this state.
\end{abstract}

\pacs{12.39.Mk, 12.39.-x}

\keywords{X(1600), Tetraquark, QCD sum rule}

\maketitle

\pagenumbering{arabic}

\section{Introduction}\label{sec1}

There has been important progress in the experimental search of
multiquark hadrons since last year. LEPS collaboration reported
evidence of the $\Theta^+$ pentaquark with $S=+1, B=+1$ and the
minimum quark content $uudd\bar s$ \cite{leps}. Such a state is
clearly beyond the conventional quark model if it is established
by future experiments. A recent review of pentaquarks can be found
in Ref. \cite{ijmpa}.

BABAR, CLEO and BELLE collaborations observed two narrow
charm-strange mesons $D_{sJ}(2317), D_{sJ}(2457)$ below threshold
\cite{babar}. These states are 160 MeV below quark model
predictions. Some authors speculate they are four-quark states
\cite{cheng,rev1}. These states may admit a small portion of $D K$
or $D^\ast K$ continuum contribution in their wave functions. But
the dominant component of $D_{sJ}(2317), D_{sJ}(2457)$ should be
$c\bar s$ \cite{dai}. This topic is reviewed in Ref. \cite{rev2}.

BELLE collaboration discovered a new narrow charmonium-like state
X(3872) in the $J/\psi \pi^+\pi^-$ channel \cite{x}. Its mass is
very close to $D{\bar D}^\ast$ threshold. Its production rate is
comparable to those of other excited charmonium states. Very
recently BELLE collaboration observed the same signal in the
$J/\psi$ $\omega$ channel \cite{x1}. Its mass and decay pattern
seems to favor its interpretation as a deuteron-like $D {\bar
D}^\ast$ molecule \cite{swan}. Naively, one would expect the lower
production rate for $D {\bar D}^\ast$ molecules.

This year SELEX collaboration reported a narrow state
$D_{sJ}(2632)$ above threshold \cite{selex}. Its dominant decay
mode is $D_s\eta$. Such an anomalous decay pattern strongly
indicates $D_{sJ}(2632)$ is a four quark state in the $SU(3)_F$
${\bf 15}$ representation with the quark content ${1\over
2\sqrt{2}}(ds\bar{d}+sd\bar{d}+su\bar{u}+us\bar{u}-2ss\bar{s})\bar{c}$
\cite{liu}. Other possible interpretations were discussed in Refs.
\cite{liu2}.

In the light meson sector, $f_0(980)/a_0(980)$ lies 10 MeV below
the $K^+K^-$ threshold. It's difficult to find a suitable position
for them within the framework of quark model. So they were
postulated to be candidates of kaon molecule or four quark states.
Recently there has accumulated some evidence of the four-quark
interpretation of the low lying scalar mesons from lattice QCD
calculation \cite{lattice1,lattice2}.

The study of multiquark states started nearly three decades ago in
the MIT bag model \cite{Jaffe,kfl}. The reaction
$\gamma\gamma\rightarrow\rho\rho$ was suggested in the search of
$q\bar qq\bar q$ resonances. Later ARGUS collaboration found
evidence of a four-quark state in the dominant partial wave
$J^PJ_z=2^+2$ in the reaction
$\gamma\gamma\rightarrow\rho^0\rho^0\rightarrow4\pi$
\cite{Augus2}. Now the observed signal was named as $X(1600\pm
100)$ with the quantum numbers $J^{PC}I^G=2^{++}2^+$ \cite{list}.

In this work, we employ QCD finite energy sum rules (FESR) to
explore whether there exists a resonance in the
$J^{PC}I^G=2^{++}2^+$ channel.

\section{Formalism}\label{sec2}

QCD sum rule approach has proven useful in extracting the masses
of the ground state hadrons \cite{shifman}. QSR approach can yield
the absolute mass scale as the lattice QCD formalism. First one
starts from the correlation function composed of the interpolating
current which strongly couples to the hadron which one wants to
study. The correlation function can be calculated using the
operator product expansion (OPE) technique. As one approaches the
resonance region from large $Q^2$, the nonperturbative power
corrections become important gradually. One gets the spectral
density of the correlation function in terms of quark and gluon
condensates at the quark level. The hadron mass enters the
spectral density of the correlation function at the hadron level.
With the quark hadron duality assumption, one can extract the
hadron mass.

The construction of a suitable interpolating current is crucial.
There are only two independent color structures for a tetraquark.
Let's first focus on a pair of $q\bar q$. Since ${\bar 3}_c \times
3_c =1_c+ 8_c$, there are only two ways to form a color singlet
tetraquark from two pairs of $\bar q q$. Both pairs are either in
the color-singlet or color-octet state simultaneously. We denote
the corresponding interpolating currents as $\eta^1, \eta^8$. If
we focus on the two quarks, $3_c \times 3_c ={\bar 3}_c+ 6_c$. The
anti-quark pair must be in either $3_c$ or ${\bar 6}_c$ color
state in order to get a color-singlet tetraquark. We denote the
corresponding interpolating currents as $\eta^{\bar 3}, \eta^6$.
It's important to note that $\eta^{\bar 3}, \eta^6$ are linear
combinations of $\eta^1, \eta^8$. We refer the reader to the
appendix for details.

We use a general interpolating current for X(1600) which is the
linear combination of $\eta^{\bar 3}$ and $\eta^6$.
\begin{eqnarray}\label{general}
 \eta_{\mu\nu}(x)&=& \eta_{\mu\nu}^{\mathbf{\bar 3}} +Y
 \eta_{\mu\nu}^{\mathbf{6}}\\
 &=&(Y+1)\bar d^l (x) \gamma_\mu u^l (x)
 \bar d^m (x) \gamma_\nu u^m(x)\nonumber +(Y-1)\bar d^l (x) \gamma_\mu u^m (x)
 \bar d^m (x) \gamma_\nu u^l(x)\nonumber + \left( g_{\mu\nu}
 \mbox{terms}\right)
\end{eqnarray}
where $Y=a+bi$ is a complex number. a and b are real numbers. The
decay constant $f_X$ for X(1600) is defined as
\begin{equation}
\langle 0| \eta_{\mu\nu} (0)| X (1600)\rangle = f_X
\varepsilon_{\mu\nu}
\end{equation}
where $\varepsilon_{\mu\nu}$ is the polarization tensor of X(1600)
meson.

We consider the following correlation function
\begin{equation}\label{cor}
i\int d^4xe^{-ipx}<0|T\{\eta_{\mu\nu}(x)
\eta^{+}_{\alpha\beta}(0)\}|0>= \Delta_{\mu\nu; \alpha\beta}(p)
\Pi(p^2) + \cdots
\end{equation}
where
\begin{equation}
\Delta_{\mu\nu; \alpha\beta}(p)= \frac
{1}{2}\left(\Delta_{\mu\alpha}(p)\Delta_{\nu\beta}(p)
+\Delta_{\mu\beta}(p)\Delta_{\nu\alpha}(p)- \frac
{2}{3}\Delta_{\mu\nu}(p)\Delta_{\alpha\beta}(p)\right)\;,
\end{equation}
\begin{equation}
\Delta_{\mu\nu}(p)=g_{\mu\nu}-p_{\mu}p_{\nu}/p^2 \;.
\end{equation}
In Eq. (\ref{cor}), we have kept the unique tensor structure for
$J^{PC}=2^{++}$ mesons: $\Delta_{\mu\nu; \alpha\beta}(p)$. We note
in passing that the $q\bar q$ meson states with $J^{PC}=2^{++}$
have been studied in \cite{reinders,shifman1}, where L=1 orbital
excitation has to be introduced. For any of its four Lorentz
indices, $\Delta_{\mu\nu; \alpha\beta}(p)$ satisfies:
\begin{equation}
q^\mu\Delta_{\mu\nu; \alpha\beta}(p)=0
\end{equation}
\begin{equation}
\Delta^{\mu}_{\mu; \alpha\beta}(p)=0 \; .
\end{equation}
The non-resonant $\rho^+\rho^+$ intermediate states do not
contribute to this tensor structure.

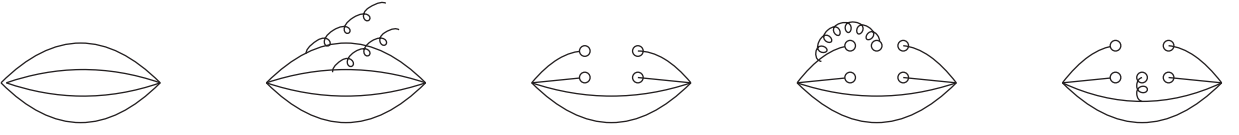
\begin{figure}[hbt]\label{diagram}
\begin{center}
\begin{picture}(500,130)(0,0)
\Curve{(25,50)(55,65)(85,50)}
 \Curve{(27,50)(55,55)(85,50)}
\Curve{(27,50)(55,45)(85,50)}
 \Curve{(25,50)(55,35)(85,50)}
\Curve{(125,50)(155,65)(185,50)}
 \Gluon(140,61)(170,80){2}{3}
 \Curve{(125,50)(155,55)(185,50)}
 \Gluon(150,54)(175,70){2}{3}
\Curve{(125,50)(155,45)(185,50)}
 \Curve{(125,50)(155,35)(185,50)}
\Curve{(225,50)(235,51)(245,52)}
\BCirc(245,52){2}
\BCirc(265,52){2}
\Curve{(265,52)(275,51)(285,50)}
\Curve{(225,50)(235,59)(245,62)}
\BCirc(245,62){2}
\BCirc(265,62){2}
\Curve{(265,62)(275,59)(285,50)}
\Curve{(225,50)(255,45)(285,50)}
 \Curve{(225,50)(255,35)(285,50)}
\Curve{(325,50)(335,60)(345,64)}
\BCirc(345,64){2}
\BCirc(365,64){2}
\Curve{(365,64)(375,60)(385,50)}
\Curve{(325,50)(335,51)(345,52)}
\BCirc(345,52){2}
\BCirc(365,52){2}
\Curve{(365,52)(375,51)(385,50)}
 \GlueArc(345,60)(11,22,190){2}{6}
\BCirc(355,64){2} \Curve{(325,50)(355,45)(385,50)}
 \Curve{(325,50)(355,35)(385,50)}
\Curve{(425,50)(435,60)(445,64)}\BCirc(445,64){2}
\BCirc(465,64){2} \Curve{(465,64)(475,60)(485,50)}
\Curve{(425,50)(435,51)(445,52)} \BCirc(445,52){2}
\BCirc(465,52){2} \Curve{(465,52)(475,51)(485,50)}
\BCirc(455,52){2} \Gluon(455,43)(455,52){2}{1}
\Curve{(425,50)(455,43)(485,50)} \Curve{(425,50)(455,35)(485,50)}
\end{picture}
\end{center}
\label{fig1} \caption{The Feynman diagrams in the calculation of
the two-point correlation function in Eq. (\ref{cor}).}
\end{figure}

The scalar complex function $\Pi (p^2)$ satisfies the following
dispersion relation
\begin{equation}\label{disp}
\Pi (p^2)=  \int {\rho(s) \over s-p^2 -i \epsilon} ds
\end{equation}
where $\rho (s)$ is the spectral density. For a narrow resonance
\begin{equation}
\rho (s)= f_X^2 \delta (s-M_X^2) + \mbox{higher states}\;.
\end{equation}

At the quark gluon level the correlation function (\ref{cor})
reads
\begin{eqnarray}\label{x}\nonumber
 i\int d^4xe^{-ipx}\{|Y+1|^2[{\bf
\mbox{Tr}}[\gamma_{\mu}S^{jn}(x)\gamma_{\alpha}S^{mi}(-x)]\times
{\bf \mbox{Tr}}[\gamma_{\nu}S^{im}(x)\gamma_{\beta}S^{nj}(-x)]\\
\nonumber
-{\bf\mbox{Tr}}[\gamma_{\mu}S^{jn}(x)\gamma_{\alpha}S^{mj}(-x)
\gamma_{\nu}S^{im}(x)\gamma_{\beta}S^{ni}(-x)]\\
\nonumber +2(|Y|^2-1)[{\bf
\mbox{Tr}}[\gamma_{\mu}S^{jm}(x)\gamma_{\alpha}S^{mi}(-x)]\times
{\bf \mbox{Tr}}[\gamma_{\nu}S^{in}(x)\gamma_{\beta}S^{nj}(-x)]\\
\nonumber
-{\bf\mbox{Tr}}[\gamma_{\mu}S^{jm}(x)\gamma_{\alpha}S^{mj}(-x)
\gamma_{\nu}S^{in}(x)\gamma_{\beta}S^{ni}(-x)]\\
\nonumber +|Y-1|^2[{\bf
\mbox{Tr}}[\gamma_{\mu}S(x)\gamma_{\alpha}S(-x)]\times {\bf
\mbox{Tr}}[\gamma_{\nu}S(x)\gamma_{\beta}S(-x)]\\
-{\bf\mbox{Tr}}[\gamma_{\mu}S(x)\gamma_{\alpha}S(-x)\gamma_{\nu}S(x)
\gamma_{\beta}S(-x)]+(\alpha\leftrightarrow\beta)]
\end{eqnarray}
where $iS(x)=\langle 0|T\{q(x) \bar q(0)\}|0\rangle $ is the full
quark propagator in the coordinate space. The $\bf\mbox{Tr}$
denotes the summation of both the color and Lorentz indices.
Throughout our calculation, we assume the up and down quarks are
massless. The first few terms of quark propagator is
\begin{equation}
iS^{ab}(x)=\frac{i\delta^ab}{2\pi^2x^4}\hat{x}+\frac{i}{32\pi^2}\frac{\lambda^n_{ab}}{2}g_cG^n_{\mu\nu}
\frac{1}{x^2}(\sigma^{\mu\nu}\hat{x}+\hat{x}\sigma^{\mu\nu})-\frac{\delta^{ab}}{12}\langle\bar
qq\rangle+\frac{\delta^{ab}x^2}{192}\langle g_s\bar q\sigma
Gq\rangle+\cdots
\end{equation}
The relevant terms that contribute to this correlator are
represented pictorially in Figure 1.

After making Fourier transformation to $\Pi(x)$ we arrive at
$\Pi(p^2)$. From the imaginary part of $\Pi(p^2)$ we extract the
spectral density $\rho(s)$.
\begin{eqnarray}\nonumber
\rho(s) &=& {1\over 2^{12}\cdot 7\cdot
\pi^6}(1+\frac{2}{3}|Y|^2)s^4  -{1\over 2^{14}\cdot 15 \cdot
\pi^6}(23|Y|^2-21(Y+Y^{\ast})+47)s^2\langle g_s^2GG\rangle  \\
 && + {5\over 18\cdot \pi ^2}(1-\frac{2}{5}|Y|^2)s
{\langle\bar qq\rangle}^2  -\frac{1}{144\cdot
\pi^2}(13|Y|^2-3(Y+Y^{\ast})-33){\langle\bar qq\rangle}{\langle
g_s\bar q\sigma G q\rangle}
\end{eqnarray}
where we have used the factorization approximation for the
high-dimension quark condensates.

\section{FESR And Numerical Analysis}\label{sec3}

In the Borel sum rule (BSR) analysis there are two parameters: the
continuum threshold $s_0$ and Borel mass $M_B$. FESR contains a
single parameter $s_0$ \cite{fesr}. The dimension of the
interpolating currents of the conventional hadrons like rho and
nucleon is not high. Both BSR and FESR yield roughly the same
results. If the interpolating current is of high dimension, the
working window of $M_B$ does not exist sometimes. In this case,
FESR may have some advantage over BSR \cite{narison}.

With the spectral density, the $n$th moment of FESR is defined as
\begin{equation}
W(n,s_0)=\int_{0}^{s_0}ds s^n \rho(s)
\end{equation}
where $n\ge 0$. With the quark hadron duality assumption we get
the finite energy sum rule
\begin{equation}\label{12}
W(n,s_0)|_{Hadron} =W(n,s_0)|_{QCD}\; .
\end{equation}
The mass can be obtained as
\begin{equation}
M^2={W(n+1,s_0)\over W(n,s_0)} \;.
\end{equation}

In principle, one can extract the threshold self-consistently from
the requirement that the hadron mass has the least dependence on
$s_0$, i.e., $ {d M^2\over d s_0} =0$. However, the threshold
value $s_0$ extracted this way is not necessarily physical
sometimes. The weight function of FESR enhances the continuum part
even more than the weight function $e^{-s/M_B^2}$ in the Borel sum
rule. One must make sure that only the lowest pole contributes to
the FESR below $s_0$. Otherwise the result will be very
misleading. To be more specific, a {\sl naive} stability region in
$s_0$ is no guarantee of a {\sl physically reasonable} value for
$s_0$. For example, the FESR with an extracted threshold
$s_0\approx 10$ GeV$^2$ is certainly irrelevant for the possible
X(1600) state.

Besides the weak dependence on $s_0$, we also require (1) the
zeroth moment $W(0, s_0) >0$ and (2) the convergence of the
operator product expansion (OPE). Higher dimension condensates
should be suppressed in order to ensure a reliable FESR based on
the converging OPE series. This convergence requirement is not
easily satisfied if the interpolating current is of high
dimension. As shown in the appendix, neither $\eta^{\bar 3}$ nor
$\eta^6$ leads to a converging FESR for a reasonable value of
$s_0$. Only with a mixed current which is the linear combination
of $\eta^{\bar 3}$ and $\eta^6$, can we make the OPE series of
FESR converging.

We use the following values of condensates: $\langle\bar
qq\rangle=-(0.24 \mbox{GeV})^3,\langle g_s^2GG\rangle =(0.48\pm
0.14)\mbox{GeV}^4,\langle g_s\bar q\sigma G
q\rangle=-m_0^2\times\langle\bar qq\rangle $,
$m_0^2=(0.8\pm0.2)$GeV$^2$. To simply numerical analysis, we
introduce variables $\epsilon_1, \epsilon_2$ which are defined as:
\begin{eqnarray}\nonumber
a^2+b^2={5\over 2}(1-\epsilon_1) \\ \nonumber a=-{1\over
12}(1+\epsilon_2) \; .
\end{eqnarray}
They satisfy two constraints: $\epsilon_1\le 1, |1+\epsilon_2|\le
6\sqrt{10(1-\epsilon_1)}$. Now the spectral density reads
\begin{eqnarray}\label{ddd}\nonumber
\rho(s) &=& {1\over 2^{12}\cdot 21\cdot \pi^6}(8-5\epsilon_1)s^4
-{1\over 2^{15}\cdot 15 \cdot
\pi^6}(216-115\epsilon_1+7\epsilon_2+7)s^2\langle g_s^2GG\rangle  \\
 && + {5\over 18\cdot \pi ^2}\epsilon_1 s
{\langle\bar qq\rangle}^2  +\frac{1}{288\cdot
\pi^2}(65\epsilon_1-\epsilon_2){\langle\bar qq\rangle}{\langle
g_s\bar q\sigma G q\rangle} \; .
\end{eqnarray}

From Eq. (\ref{ddd}) it's clear that we can adjust the variables
$\epsilon_1, \epsilon_2$ to suppress $D=6, 8$ condensates. In fact
there exists a parameter space of $(\epsilon_1, \epsilon_2)$, with
which there exists a stable FESR plateau in the $M_X$ {\sl vs}
$s_0$ curve. The extracted $s_0$ at the stable plateau is
physically reasonable. With this $s_0$ the zeroth FESR moment is
positive. and the OPE is convergent. One typical set of these
variables is $\epsilon_1=-0.02, \epsilon_2=-2$. After we divide
each piece in the FESR by the perturbative term, the zeroth moment
reads
\begin{equation}
W(0,s_0)\sim 1-{3.53\over s^2_0} -{2.79\over s_0^3}-{1.95\over
s^4_0}\; .
\end{equation}
The positivity requirement leads to $s_0>2.3$ GeV$^2$.

\begin{figure}[hbt]
\begin{center}
\scalebox{0.8}{\includegraphics{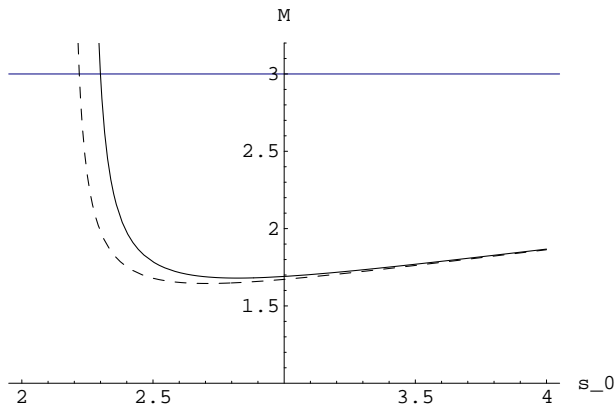}}
\end{center}\label{bi0m}
\caption{The variation of the mass $M$ with the threshold $s_0$
(in unit of $\mbox{GeV}^2$) for the current (\ref{general}). The
long-dashed and solid curves correspond to FESR when truncated at
dimension $D=6$ and $D=8$ respectively.}
\end{figure}

For $s_0>2.3$ GeV$^2$, the OPE also converges well. The variation
of $M_X$ with $s_0$ is presented in Figure 2. The extracted mass
is
\begin{equation}
M_X=(1.65\pm 0.15) \mbox{GeV} \;.
\end{equation}
As can be seen from Figure 2, the value of $M_X$ does not change
much if we truncate the OPE at $D=6$. This is another sign of
convergence of our FESR.

\section{Discussion}\label{sec4}

The state $X(1600)$ with $J^{PC}I^G=2^{++}2^+$ has inspired many
theoretical papers \cite{russia}. The potential scattering of
$\rho^0\rho^0$ via the $\sigma$ meson exchange was proposed to
explain the signal in Ref. \cite{bajc}. In the diquark cluster
model \cite{cluster}, the mass and decay width of $X(1600)$ is
studied together with other tetraquarks assuming the $\mathbf{\bar
3}_c$ color wave function for two quarks. Using the potential
model $X(1600)$ mass was estimated to be $1544$ MeV recently
\cite{vijande2}.

In this work we have investigated this state using QCD finite
energy sum rule. With any single current $\eta^1$, $\eta^8$,
$\eta^{\bar 3}$, $\eta^6$, the operator product expansion of the
corresponding correlator does not converge. After we consider the
linear combination of $\eta^{\bar 3}$ and $\eta^6$, we have
derived a converging FESR and observe a resonance signal at 1.65
GeV, which is close to the experimentally observed X(1600) state.
Our analysis indicates both the "hidden color" and coupled channel
effects may be important in the multiquark system.

From charge, angular momentum, parity, isospin, C parity, and G
parity conservation, we may obtain the allowed decay modes of
$X(1600)$. G parity requires even number of pions in the final
state for pure pion final states. For example, $X^0(1600) \to
\rho^0 \pi^0, \omega \pi^+\pi^-$ is forbidden by G parity. C
parity forbids $X^0(1600) \to \omega (3\pi^0)$. Isospin
conservation forbids the following decay modes: $X^0(1600) \to
\omega \pi^0, \omega\omega$. The possible modes are $X^0(1600) \to
\pi^0 \pi^0, \pi^+\pi^-, 4\pi, \rho^+\rho^-, \rho^0\rho^0$.
Angular momentum conservation requires D-wave decay for the two
pion mode. Therefore, two pion decay modes may be suppressed
compared with S-wave $\rho\rho$ mode.

Filippi et al. reported the possible existence of the
$J^{PC}I^G=0^{++}2^+$ state decaying into $\pi^+\pi^+$ in the
reaction $\bar np\longrightarrow \pi^+\pi^+\pi^-$ \cite{fili}. One
the other hand, some constraint has been obtained on the possible
resonance in the scalar isotensor channel from the phase shift
analysis of $\pi\pi \to \rho\rho\to \pi\pi$ scattering \cite{zou}.
Constraint on the D-wave I=2 resonance from the similar analysis
of the same reactions will be very desirable.

The process $\gamma \gamma\to \rho\rho\to X(1600)\to 4\pi$ was
suggested and used in the search of $X(1600)$. We suggest the
following reactions to look for this charming state.
\begin{itemize}
\item $J/\psi$ decays at BES, CLEO and $\Upsilon$ decays at BELLE,
BABAR

Symmetry considerations requires $X(1600)$ is produced together
with odd number of pions in $J/\psi$ decays. One pion mode is
forbidden by isospin conservation. So the favorable decay chain is
$J/\psi \to X(1600) +3\pi \to 2\rho+3\pi \to 7\pi$. There is also
some chance in $J/\psi \to X(1600) +3\pi \to 2\pi+3\pi \to 5\pi$.

\item Hadron reactions especially charge exchange processes

Charge exchange processes are useful for the production of
$X^{++}(1600)$ if such a state exists. Some of them are $\pi^+ N
\to \pi^- + X^{++}(1600) + N \to 5\pi +N$, $ p + N \to n +
N^\prime + X^{++}(1600)$ where $N$ is either a nucleon or nuclei.
The later process can be studied at CSR facility at Lan Zhou.

\item Anti-proton annihilation on the proton or deuteron targets

Let's take $X(1600)$ production from anti-proton annihilation on
the proton target as an example. Since the isospin of $p \bar p$
is either 0 or 1, $X(1600)$ should be accompanied by one or
several pions. However, kinematics allows only one pion decay
mode: ${\bar p} + p \to X(1600) + \pi^0 \to 5\pi$. So we get
$C_{p\bar p}=(-)^{L_{p\bar p}+S_{p\bar p}}=+, I_{p\bar p}=1$.

(1) When the $p \bar p$ pair is in the S-wave, $L_{p\bar
p}=S_{p\bar p}=J_{p\bar p}=0, P_{p\bar p}=-$. Angular momentum and
parity conservation requires $L_{\pi X(1600)}=2$; (2) When the $p
\bar p$ pair is in the P-wave, $L_{p\bar p}=S_{p\bar p}=1$,
$P_{p\bar p}=+$, $J_{p\bar p}=1$ or 2. We get $L_{\pi X(1600)}=1$;
(3) When the $p \bar p$ pair is in the D-wave, $L_{p\bar p}=2,
S_{p\bar p}=0, J_{p\bar p}=2, P_{p\bar p}=-$. We have $L_{\pi
X(1600)}=0$.
\end{itemize}

\section*{Acknowledgments}

S.L.Z. thanks B.-S. Zou and H.-Q. Zheng for helpful discussions.
This project was supported by the National Natural Science
Foundation of China under Grants 10375003 and 10421003, Ministry
of Education of China, FANEDD, Key Grant Project of Chinese
Ministry of Education (NO 305001) and SRF for ROCS, SEM.

\section{Appendix}

In this appendix we list the currents with different color
structure and their spectral densities.

The color-singlet $\rho^+ (x) \rho^+(x)$ type interpolating
current reads
\begin{equation}\label{current1}
 \eta^1_{\mu\nu} (x) =\bar d^i (x)\gamma_\mu u^i (x)\bar d^j (x)
 \gamma_\nu u^j(x)   -1/4 g_{\mu\nu}\bar d^i(x) \gamma_\rho u^i(x)
 \bar d^j(x)\gamma^\rho u^j (x) \;,
\end{equation}
which "i, j" are the color indices.

The current with color octet structure reads
\begin{equation}\label{current2}
 \eta^8_{\mu\nu} (x) =\bar d^i (x)\left({\lambda^a\over 2}\right)_{ij}
 \gamma_\mu u^j (x)\bar d^m (x) \left({\lambda^a\over 2}\right)_{mn}
 \gamma_\nu u^n(x)-{1\over 4} g_{\mu\nu}\bar d^i(x)\left({\lambda^a\over 2}
 \right)_{ij}\gamma_\rho u^j(x)\bar d^m(x) \left({\lambda^a\over2}
 \right)_{mn}\gamma^\rho u^n (x)
\end{equation}
where ${\lambda^a\over 2}$ is the $SU(3)_c$ generator.

Using the identity of the $\lambda$ matrix
\begin{equation}
\sum\limits_{a}\frac{\lambda_{ij}^a}{2}\cdot\frac{\lambda_{kl}^a}{2}
=\frac{1}{2}(\delta_{il}\delta_{jk}-\frac{1}{3}\delta_{ij}\delta_{kl})
\end{equation}
we can rewrite Eq. (\ref{current2}) as
\begin{equation}
 \eta_{\mu\nu}^8 (x)
 =-\frac 16\bar d^l (x) \gamma_\mu u^l (x)\bar d^m (x) \gamma_\nu u^m(x)
 +\frac 12\bar d^l (x) \gamma_\mu u^m (x)\bar d^m (x) \gamma_\nu u^l(x)
+ \left(g_{\mu\nu} \mbox{terms}\right) \; .
\end{equation}

Alternatively, when two quarks are in the $\mathbf{\bar 3}_c$
state, we have
\begin{equation}\label{diquark-type}
 \eta_{\mu\nu}^{\mathbf{\bar 3}_c} (x)=\bar d^l (x) \gamma_\mu u^l (x)
 \bar d^m (x) \gamma_\nu u^m(x)\nonumber -\bar d^l (x) \gamma_\mu u^m (x)
 \bar d^m (x) \gamma_\nu u^l(x)\nonumber + \left( g_{\mu\nu}
 \mbox{terms}\right) \; .
\end{equation}
The current with the $\mathbf{6_c}$ color structure is
\begin{equation}\label{6c-type}
 \eta_{\mu\nu}^{\mathbf{6_c}} (x)=\bar d^l (x) \gamma_\mu u^l (x)
 \bar d^m (x) \gamma_\nu u^m(x)\nonumber +\bar d^l (x) \gamma_\mu u^m (x)
 \bar d^m (x) \gamma_\nu u^l(x)\nonumber + \left( g_{\mu\nu}
 \mbox{terms}\right)\; .
\end{equation}

The spectral densities of the above four currents
\begin{equation} \rho^1 (s)={5\over 2^{14}\cdot
21\cdot \pi^6}s^4  -{7\over 2^{12}\cdot 15\cdot \pi ^6}s^2\langle
g_s^2GG\rangle+ {1\over 24\pi ^2}s {\langle\bar
qq\rangle}^2+\frac{7}{288 \cdot \pi^2}{\langle\bar
qq\rangle}{\langle g_s\bar q\sigma G q\rangle}\; .
\end{equation}
\begin{equation} \rho^8 (s)={1\over 2^{13}\cdot
27\cdot \pi^6}s^4  -{127\over 2^{16}\cdot 135 \cdot \pi
^6}s^2\langle g_s^2GG\rangle+ {1\over 36\cdot \pi ^2}s
{\langle\bar qq\rangle}^2+\frac{131}{64 \cdot 81\cdot
\pi^2}{\langle\bar qq\rangle}{\langle g_s\bar q\sigma G
q\rangle}\; .
\end{equation}
\begin{equation} \rho^{\mathbf{\bar 3}_c} (s)={1\over 2^{12}\cdot
7\cdot \pi^6}s^4  -{47\over 2^{14}\cdot 15 \cdot \pi ^6}s^2\langle
g_s^2GG\rangle+ {5\over 18\cdot \pi ^2}s {\langle\bar
qq\rangle}^2+\frac{11}{48\cdot \pi^2}{\langle\bar
qq\rangle}{\langle g_s\bar q\sigma G q\rangle} \; .
\end{equation}
\begin{equation} \rho^{\mathbf{6}_c} (s)={1\over 2^{11}\cdot
21\cdot \pi^6}s^4  -{23\over 2^{14}\cdot 15 \cdot \pi
^6}s^2\langle g_s^2GG\rangle- {1\over 9\cdot \pi ^2}s {\langle\bar
qq\rangle}^2-\frac{13}{144\cdot \pi^2}{\langle\bar
qq\rangle}{\langle g_s\bar q\sigma G q\rangle} \; .
\end{equation}

In order to study the convergence of OPE series in FESR, we divide
each piece in the zeroth moments by the four quark condensate.
\begin{eqnarray}
W^{1}(0,s_0)\sim 7.5\times 10^{-3} s^3_0-4.7\times 10^{-2} s_0
+1-{0.9\over s_0}\; .\\ \nonumber W^{8}(0,s_0)\sim 3.5\times
10^{-3}s^3_0+8.9\times 10^{-3} s_0 +1-{1.5\over s_0}\; .\\
\nonumber W^{\mathbf{\bar 3}}(0,s_0)\sim 2.7\times 10^{-3}
s^3_0+1.2\times 10^{-2} s_0 +1-{1.3\over s_0}\; .\\ \nonumber
W^{\mathbf{6}}(0,s_0)\sim -9.0\times 10^{-3} s^3_0+2.9\times
10^{-2} s_0 +1-{2.6\over s_0}\; .
\end{eqnarray}
Clearly these moments converge only when the threshold parameter
is very large, $s_0>10$ GeV$^2$. Such a large $s_0$ is irrelevant
for X(1600).

\end{document}